# Condensation signatures of photogenerated interlayer excitons in a van der Waals heterostack

**Authors:** Lukas Sigl[1], Florian Sigger[1], Fabian Kronowetter[1], Jonas Kiemle[1], Julian Klein[1], Kenji Watanabe[2], Takashi Taniguchi[2], Jonathan J. Finley[1,3], Ursula Wurstbauer[1,4*] and Alexander W. Holleitner[1,3*]

[1] Walter Schottky Institute and Physics Department, TU Munich, Am Coulombwall 4a, 85748 Garching, Germany.
[2] National Institute for Materials Science, Tsukuba, Ibaraki 305-0044, Japan.
[3] Munich Center for Quantum Science and Technology (MCQST), Schellingstr. 4, 80799 München, Germany
[4] Institute of Physics, Westfälische Wilhelms-Universität Münster, Wilhelm-Klemm-Str.10, 48149 Münster, Germany.

**Atomistic van der Waals heterostacks are ideal systems for high-temperature exciton condensation because of large exciton binding energies and long lifetimes. Charge transport and electron energy-loss spectroscopy showed first evidence of excitonic many-body states in such two-dimensional materials. Pure optical studies, the most obvious way to access the phase diagram of photogenerated excitons have been elusive. We observe several criticalities in photogenerated exciton ensembles hosted in MoSe$_2$–WSe$_2$ heterostacks with respect to photoluminescence intensity, linewidth, and temporal coherence pointing towards the transition to a coherent quantum state. For this state, the occupation is 100% and the exciton diffusion length is increased. The phenomena survive above 10 kelvin, consistent with the predicted critical condensation temperature. Our study provides a first phase-diagram of many-body interlayer exciton states including Bose Einstein condensation.**

Increasing the interaction strength between quasiparticles can cause strong correlations, collective phenomena and even the transition to new emergent quantum phases of the many-body state that are substantionally different from properties of the weakly interacting system. Many examples exist in condensend matter systems where interacting electrons, spins and phonons form novel many-body ground states including superconductivity, superfluidity, charge-density wave phases, polariton condensates and spin-liquids[1–4]. Another predicted many-body state of quasiparticles is a Bose-Einstein condensate of interacting excitons[5,6], that are bound electron-hole pairs in semiconductors. Their many-body phase diagram predicts quantum degeneracy and Bose-Einstein condensation at reduced temperatures and sufficiently high exciton densities[7–11]. In transport experiments, condensed phases of excitons have been reported in quantum Hall states for both graphene and coupled semiconductor quantum wells[12,13]. The weak exciton binding in these systems, together with the presence of low-energy Goldstone modes makes it difficult to access the condensates optically and limits the condensation temperature to about 1 K or lower[5]. Atomically thin, semiconducting monolayers such as transition metal dichalcogenides (TMDCs) exhibit a large exciton binding energy (about 0.5 eV)[14] and the possibility to form van der Waals heterostacks[15,16]. In turn, they provide a promising solid state platform for exploring high-temperature exciton condensation and realizing condensate-based applications on a chip[9,17,18]. To diminish the limitations, such as short radiative lifetimes[19] and biexcitons[20] present for intralayer excitons in individual monolayers, the spatial overlap of the Coulomb-bound electron hole pairs can be reduced by placing them in adjacent layers in a double-layer heterostack[21]. Long lifetimes of several tens of nanoseconds, while the binding energies remain substantial (>0.1 eV), have been demonstrated for such photogenerated interlayer excitons[21–23]. Moreover, they act like oriented electric dipoles supporting dipole-dipole interactions and inducing exciton correlations at high densities[24,25]. Theoretically, room temperature condensation may be possible for these interlayer excitons in TMDC double layers since the maximum condensation temperature is limited by exciton ionization in the high-density regime, and it has been predicted to be a fraction of the exciton binding energy[9]. Nevertheless, the exploration of the many-body phase diagram of photogenerated interlayer excitons in TMDCs including quantum degeneracy and condensation has remained elusive so far.

We present clear signatures of the condensation of photogenerated excitons in MoSe$_2$-WSe$_2$ heterostacks at elevated temperatures via several criticalities and map the many-body phase diagram of the photogenerated interlayer excitons. The device structure under investigation consists of two rotationally-aligned TMDC monolayers (Fig. 1a and supplementary material)[22]. The cryogenic photoluminescence spectra exhibit only one very sharp emission line (1) at low excitation powers, which dominates all other optical transitions including those arising from intralayer excitons $X_{1s}^{MoSe2}$, $X_{1s}^{WSe2}$ and the trions $X_T^{MoSe2}$, $X_T^{WSe2}$ in the monolayer MoSe$_2$ and WSe$_2$ (supplementary material)[26]. The emission energy of this state is about 6-7 meV below to further emission peaks named (2) and (3) in Fig. 1b that appear only for higher excitation powers. Peaks (2) and (3) have been reported before, both by our group and others[22,27–30]. We interpret them arising from reciprocally indirect transitions from spin-orbit-split states at $\Sigma$ in the conduction band (CB$^{MoSe2}$) of MoSe$_2$ to $K$ ($K$') in the valence band (VB$^{WSe2}$) of WSe$_2$, where the photon emission is linked with the emission of acoustic phonons[28]. Intriguingly, the intensity of the peak (1) is enhanced by two orders of magnitude at the lowest temperature (Fig. 1c). Since the decay time of this peak also increases by a factor of two, the enhanced photon emission suggests that an increasingly dense part of the exciton ensemble emits into the light cone (Fig. 1d)[26]. Notably, this strong enhancement of emission is not observed for peaks (2) and (3) and their decay times remain constant in temperature and two orders of magnitude below the one of the emission peak (1) (supplementary materials). Plotting the phase diagram

- temperature vs. exciton density $n$ – for peak (1), we observe that for a temperature below 8 K (open circles in Fig. 1e), the excitons are below the degeneracy limit obeying[6,9]

$$k_B \cdot T_D = 2\pi \hbar^2 n / m_X, \qquad (1)$$

with $k_B$ the Boltzmann constant, $T_D$ the degeneracy temperature, $\hbar$ the reduced Planck's constant, and $m_X = m_e + m_h$ the effective exciton mass with $m_e = 0.54$ ($m_h = 0.36$) the effective electron (hole) mass[31]. The measurements presented were performed on two devices for temperatures between 3 and 150 K during several independent cool-downs demonstrating the robustness of the experimentally observed phenomena.

Below the degeneracy temperature $T_D$, all interlayer excitons emit photons via peak (1) (Fig. 1b); i.e. the underlying exciton state exhibits a relative occupation close to 100% (Fig. 2a)[26]. Moreover, the FWHM of peak (1) reduces to a constant value less than ~5 meV below $T_D$ (Fig. 2b), although the intensity and lifetime still increase (Fig. 1c). Furthermore, we observe a blue-shift of peak (1) up to 2 meV (supplementary material) that is consistent with a repulsive dipole-dipole interaction within the interlayer exciton ensemble[24,25], particularly, since they have a longer lifetime and therefore an increasing density at the lowest temperatures (Fig. 1c bottom and Fig. 1e). Below $T_D$, the lateral exciton diffusion length related to state (1) is enhanced by a factor of 2 to a value of 1.3 µm (supplementary material). In turn, the majority of the interlayer excitons in state (1) diffuse out of the confocal laser spot (~0.5 µm) at the lowest temperatures, while most of the excitons emitting into peak (2) and (3) remain within the excitation spot before they recombine. We explain the increased diffusion length observed for state (1) as arising from the increased exciton lifetime as in Fig. 1c and note that the diffusion coefficients are constant within the investigated temperature range ($D_{(1)} = 0.6 \pm 0.5$ cm$^2$/s; $D_{(2)} = 1.3 \pm 0.5$

cm$^2$/s; $D_{(3)}$ = 0.4 ± 0.4 cm$^2$/s)[26,30]. In this respect, we do not observe a Berezinskii-Kosterlitz-Thouless (BKT) transition to a state with modified expansion properties[6].

The FWHM of peak (1) decreases as a function of exciton density, until it reaches a minimum value below ~5 meV in the degenerate regime (triangles in Fig. 3a). For comparison, the dotted line in Fig. 3a marks the degeneracy density $n_D$ as described by equation (1) for the given temperature, and we added the triangles of Fig. 3a into the phase diagram of Figure 1e. The emission energy increases by ~1 meV vs. exciton density in the degenerate regime (Fig. 3b), which is again consistent with a repulsive dipole-dipole interaction[24,25]. For this experiment, the excitation energy is chosen to be $E_{photon}$ = 1.59 eV, which is below the intralayer exciton energies in the MoSe$_2$ and WSe$_2$ monolayers to minimize their impact on the interlayer excitons[26]. The plotted density $n$ only considers interlayer excitons which recombine via the light cone (Fig. 1d). For an exciton density exceeding $n_{transition}$ ~2.5 · 10$^{11}$cm$^{-2}$, we observe that the intensity of peak (1) saturates and peaks (2) and (3) gradually occur in the spectra (supplementary material). Above this critical density, the FWHM of peak (1) rapidly increases (Fig. 3a) and the relative occupation of state (1) is less than 100%. This transition at $n_{transition}$ cannot arise from a Mott-transition, since all emission peaks have long decay times which remain entirely consistent with the photon emission from interlayer excitons (Supplementary material). For comparison, the expected theoretical value of $n_{Mott}$ for our heterostack exceeds 10$^{12}$ cm$^{-2}$ [9].

Figure 4a shows the temporal coherence visibility measured for emission peak (1) in the degenerate regime below $T_D$ demonstrating that the coherence adheres to the Wiener-Khinchine theorem[32,33]. In other words, the Fourier transform of the Lorentzian-shaped emission spectra as in Fig. 1b overlaps with the experimentally determined coherence visibility with a maximum value of (88 ± 7 %). Above the degeneracy temperature $T_D$, the temporal coherence drops significantly (Figs. 4b,c). Fitting the visibility with exponential decays (black lines in Figs. 4a-c) reveals a fast drop of the temporal coherence time $\tau_c$ and length $l_c$ above $T_D$ (Fig. 4d). Below $T_D$, the coherence time is limited to a time scale of ~300 fs, which is approximately five to six orders of magnitude faster than the decay time of peak (1) as depicted in Fig. 4d. In turn, the Lorentzian-shaped emission spectrum of state (1) as in Fig. 1b is not radiation-limited (supplementary material). Instead, the fast time scale suggests that below $T_D$, the coherence is limited by the coupling to acoustic phonons (LA, TA or ZA) during the photon emission (Fig. 1d)[34]. The photon emission process then has an effect similar to collisional broadening in atom ensembles[35] as the interaction with acoustic phonons gives rise to a phase-shift of the emitted photons (Fig. 4e). This phase shift due to phonon assisted interband recombination dominates the FWHM of the Lorentzian-shaped lines. The temperature dependence of the interlayer exciton emission energies yield an energy of $E_p$ ~ 14 meV (~295 fs) of phonons interacting with the interlayer excitons[26] (Supplementary material).

The observed energy difference between peaks (2) and (3) (~25 meV) is consistent with the spin-orbit splitting of the $CB^{MoSe2}$ ($\Delta SOI$ in Fig. 1d) and not with the one in the $VB^{WSe2}$ (~450 meV)[31,36]. A phenomenological polaron fit below 150 K gives an exciton-phonon coupling with a Huang-Rhys factor of S ~ 1.70 for peak (2) and 1.85 for peak (3) and an average energy of interacting (acoustic) phonons of $E_p$ ~ 14 meV for both peaks[28] (Supplementary material). Emission peak (1) can be detected only below 50 K with an emission energy that, unusually, first reduces below and then increases beyond $T_D$ ~ 8 K (Supplementary material). However, as discussed above, an energy scale of ~14 meV limits its temporal coherence to a time scale of ~300 fs. For comparison, the polaron fit for the direct 1s-exciton transition in $MoSe_2$ ($WSe_2$) gives S ~ 2.39 (2.28) and $E_p$ ~ 16 meV (18.7 meV). Since the Huang-Rhys factors are different to the ones of the direct transitions at $K$ ($K'$) and since they interact with acoustic phonons, we identify peaks (2) and (3) arising from reciprocally indirect transitions from SO-split states at $\Sigma$ in the $CB^{MoSe2}$ to $K$ ($K'$) in the $VB^{WSe2}$ (as sketched in Fig. 1d)[28].

We now summarize several arguments which indicate that peak (1) cannot be identified as a single-quasiparticle transition, while peaks (2) and (3) results from exciton recombination in a single particle framework. On the one hand, peak (1) could correspond to an optical transition from the minima at $K$ ($K'$) in the $CB^{MoSe_2}$ to the maxima at $K$ ($K'$) in the $VB^{WSe_2}$. However, it is known that this reciprocally direct transition should have a higher energy difference than the indirect one between $\Sigma$ in the $CB^{MoSe_2}$ to $K$ ($K'$) in the $VB^{WSe_2}$ especially at high excess charge carrier densities[22,37] which is contrary to our observations. On the other hand, peak (1) could relate to a transition from $\Sigma$ ($CB^{MoSe_2}$) to $\Gamma$ ($VB^{WSe_2}$). However, it is counterintuitive that such a reciprocally indirect state would have an increased diffusion length and strongly increasing luminescence intensity at lowest temperatures. At the same time, the corresponding phonon-density supporting this indirect single-particle transition is known to be small with negligible electron-phonon coupling[38].

Instead, we propose that the emission peak (1) corresponds to a many-body state of interlayer excitons resulting from a repulsive interaction with an energy at the few meV level, and that this many-body state is intrinsically related to the single-particle transitions of peaks (2) and (3) (cf. Fig. 1d). Again, two of the most prominent arguments for the interrelation are that peak (1) changes its general behavior as soon as peak (2) appears (dashed lines in Fig. 4f and 3), and that the exciton-phonon coupling $E_p \sim 14$ meV for peaks (2) and (3) coincidences with the collision broadening of peak (1) (~300 fs). The interrelation can be further seen in the decay time of peak (2) which starts with a rather long time of ~20 ns close to $n_{Transition}$, when peak (1) is saturated to the rather short lifetime of peak (3) of ~1 ns as soon as peak (1) is not resolvable anymore at an overall high exciton density (supplementary material).

The evidence for a many-body state is threefold. Firstly, the state has a relative occupancy of 100% for experimental conditions such that equation (1) is fulfilled. Secondly, below the degeneracy temperature $T_D$, the intensity of peak (1) is consistent with an increasingly dense exciton population emitting more and more into the light cone. Thirdly, below $T_D$, the FWHM of peak (1) is independent from thermal broadening but it decreases with increasing exciton density, and the Wiener-Khinchine theorem is fulfilled with a limiting temporal coherence of 300 fs. From the polaron fits, we know that up to ~100 K the energy (coherence time) of the interacting phonons is $E_p$ ~ 14 meV (300 fs). In turn, for temperature larger $T_D$, we can deduce that the exponential increase of the coherence stems from the excitons (cf. exponential fit in Fig. 4d). Below $T_D$, the overall emission coherence is limited by the phonons that are intrinsic to the recombination and photon emission process. Peaks (1) and (2) exhibit an energy difference of ~6-7 meV. The many-body interaction most likely results from the dipole-dipole interaction ~1-2 meV and possibly, by the layer breathing mode as is known for TMDC-heterostacks with an energy of 3.7-4.9 meV (30-40 cm$^{-1}$)[39]. In this picture, the underlying exciton state of peak (2) limits the coherence of peak (1) as soon as the FWHM of peak (1) exceeds the energy difference between the two peaks either by exciton-exciton interactions within the dense ensemble (dashed lines in Figs. 1e, 3a, and 4f) or by thermal fluctuations (dotted lines in Figs. 1e and 4d).

In conclusion, we optically generated a high-density ensemble of interlayer excitons in a 2D van der Waals heterostack. In photoluminescence studies, we have observed a threshold behavior on temperature and density, which is consistent with exciton condensation. The observations persist to a temperature between 10 and 20 K. We describe an initial experimental phase diagram of quantum degenerate interlayer excitons. Our results open up opportunities to explore the optical properties of exciton condensates and high-temperature superconductivity in exciton ensembles[9].

**Acknowledgments:** We thank R. Schmidt, R. Rappaport, and F. Dubin for discussions. We gratefully acknowledge financial support by the Deutsche Forschungsgemeinschaft (DFG) via projects WU 637/4- 1 and HO 3324/9-1 and the excellence cluster Munich Center for Quantum Science and Technology (MCQST). K.W. and T.T. acknowledge support from the Elemental Strategy Initiative conducted by the MEXT, Japan and the CREST (JPMJCR15F3), JST.

**Author contributions:** L.S., U.W. and A.W.H. conceived and designed the experiments. J.K and F.S. prepared the samples, K.W. and T.T. provided high-quality hBN bulk crystals, L.S., F.K., J.K. and F.S. performed the experiments. L.S., F.K., J.J.F., U.W. and A.W.H. analyzed and discussed the data, L.S. and A.W.H. wrote the manuscript with input from all authors.

**Competing interests:** The authors declare no competing interests.

**Data and materials availability:** The data presented in this manuscript are available from the authors upon reasonable request.

**Supplementary Materials:**

Materials and Methods, including optical images of samples, complete spectra with inter- and intralayer excitons, exciton diffusion experiments, saturation and lifetime determination, Lorentzian- and Gaussian-shaped coherence fits, and polaron fits.

Figures S1-S11

Movie S1

**Correspondence and requests for materials** should be addressed to:

wurstbauer@uni-muenster.de and holleitner@wsi.tum.de

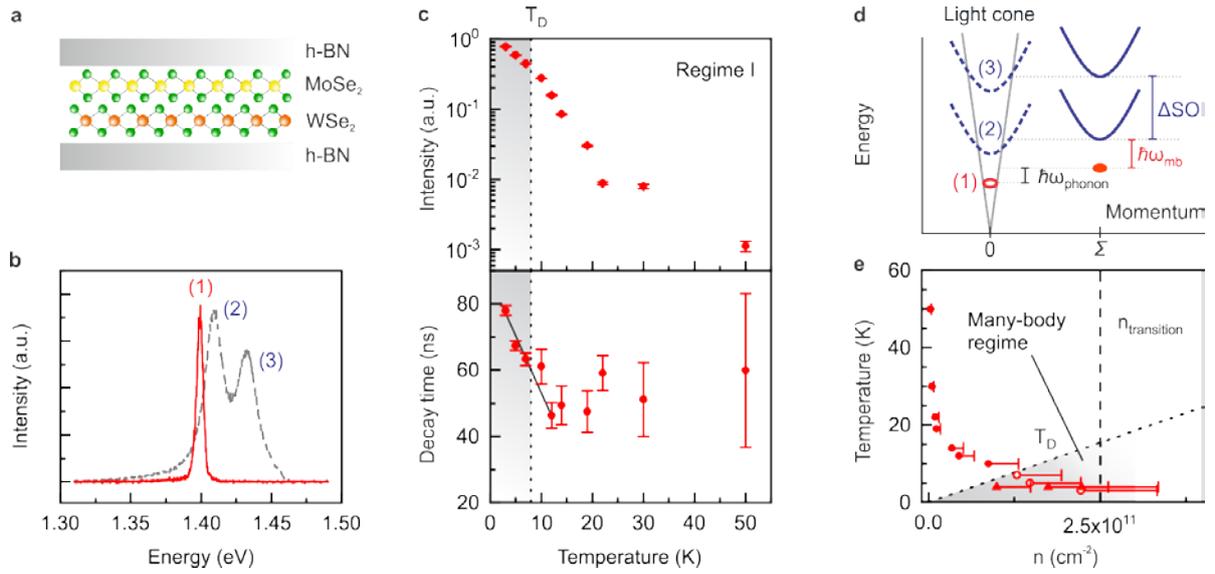

**Fig. 1 | Phase diagram of many-body interlayer exciton state. a,** Schematic of a MoSe$_2$/WSe$_2$ heterostack encapsulated in hBN. **b,** Photoluminescence spectra with many-body state – emission peak (1) – as it is observable at a low excitation power of 200 nW, and peaks (2) and (3) at 420 µW (excitation energy $E_{photon}$ = 1.946 eV and bath temperature $T$ = 4 K). All emission peaks stem from interlayer excitons. **c,** Intensity (top) and lifetime (bottom) of many-body state vs. bath temperature. Black line is a guide to the eye ($E_{photon}$ = 1.59 eV). **d,** Energy dispersion of interlayer exciton states (1), (2), and (3) with acoustic phonons at energy $\hbar\omega_{phonon}$ connecting the exciton states at a momentum of $\Sigma$ with the light cone and $\hbar\omega_{mb}$ ($\Delta$SOI) the energy of the many-body (spin orbit) interaction. **e,** Phase diagram for many-body state: temperature vs. exciton density with dotted line according to equation (1) and $T_D$ the degeneracy temperature. Dashed line for a transition density $n_{transition}$ between many-body regime and the regime where peaks (2) and (3) occur.

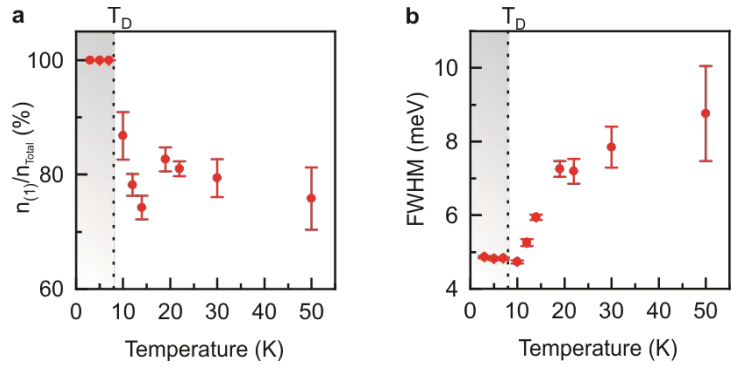

**Fig. 2 | Temperature dependence and occupation saturation of many-body state.**
**a,** Relative occupation of many-body state with respect to exciton states (2) and (3) vs. temperature. **b,** Full width at half maximum (FWHM) of many-body state vs. temperature ($E_{photon}$ = 1.59 eV).

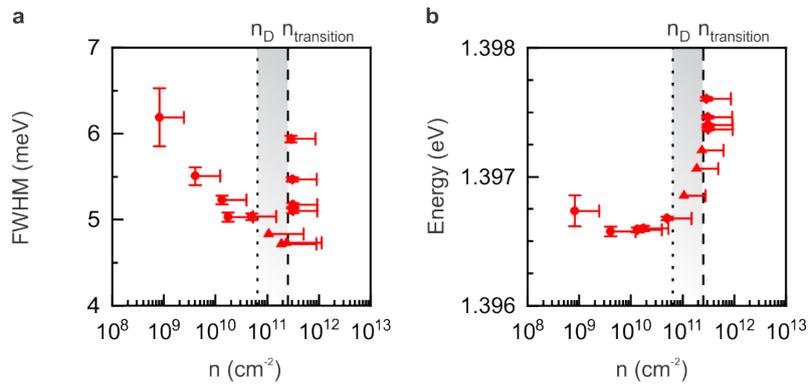

**Fig. 3 | Density dependence of many-body state and exciton-exciton interaction.**
**a,** Photoluminescence FWHM of many-body state vs. exciton density with dotted line according to equation (1) and $n_D$ the degenerate density at $T$ = 4 K. $n_{transition}$ marks transition to the regime where peaks (2) and (3) occur (dashed line). **b,** Emission energy of many-body state vs. exciton density ($E_{photon}$ = 1.59 eV).

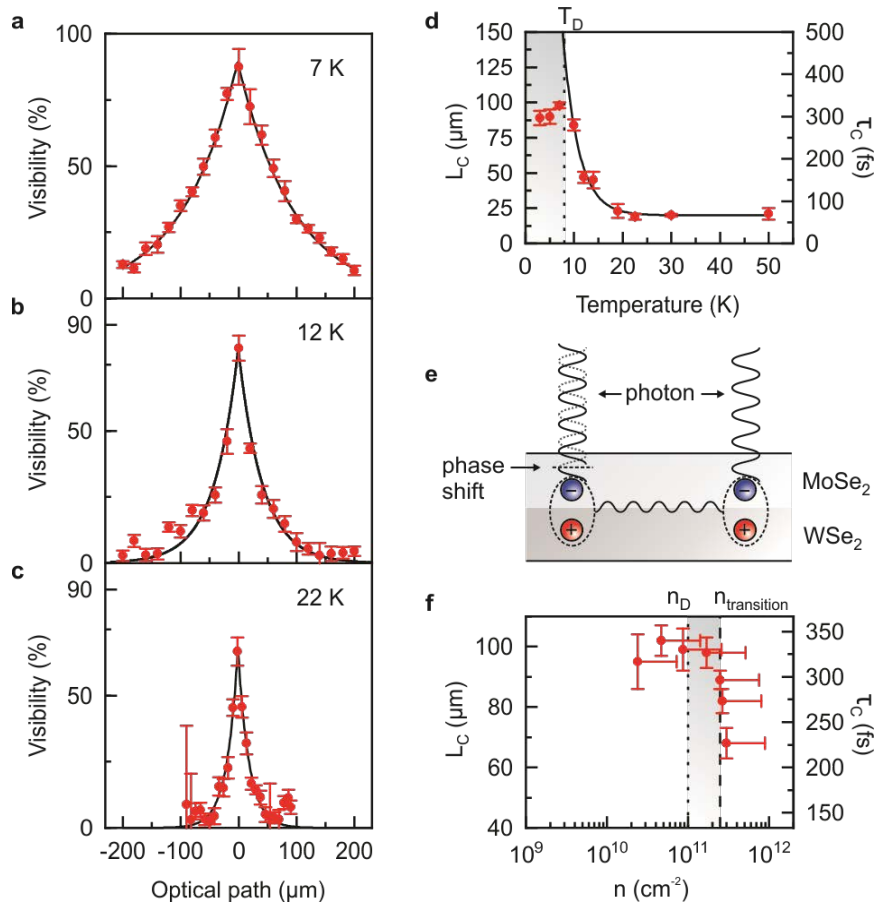

**Fig. 4 | Fourier limited temporal coherence in many-body regime. a,-c,** Temporal coherence visibility vs. optical path for many-body state for temperatures of 7 K, 12 K, and 22 K ($E_{photon}$ = 1.59 eV). Black lines are exponential fits. **d,** Temporal coherence length $l_c$ and corresponding coherence time $\tau_c$ for many-body state vs. temperature with an exponential fit above $T_D$. **e,** Sketch of photon emission with a phase-shift relative to a second emitted photon with the two related excitons interacting with each other (wavy line). **f,** Temporal coherence length vs. exciton density in the many-body state at $T$ = 7 K.

Supplementary Materials for

# Condensation signatures of photogenerated interlayer excitons in a van der Waals heterostack

Lukas Sigl, Florian Sigger, Fabian Kronowetter, Jonas Kiemle, Julian Klein, Kenji Watanabe, Takashi Taniguchi, Jonathan J. Finley, Ursula Wurstbauer*, Alexander W. Holleitner*

*Correspondence to: wurstbauer@uni-muenster.de and holleitner@wsi.tum.de

**This PDF file includes:**

    Materials and Methods
    Figs. S1 to S12

## Materials and Methods

Materials preparation

The samples were prepared using an all-dry viscoelastic stamping method. The bulk material was mechanically cleaved and subsequently transferred onto a PDMS stamp fixed on a glass slide. The bulk crystals were grown synthetically by hqgraphene ($MoSe_2$/$WSe_2$) and by K.W. and T.T. (hBN). After identification by optical contrast, the flakes were transferred from the stamp onto the $SiO_2$ substrate in a home-built heatable stage. Clean interfaces were achieved by wet chemical cleaning and annealing in between every stacking step of the heterostructure. A built-in rotation stage facilitated alignment of the flake edges to a precision of ±1° in order to achieve a commensurate alignment of the crystal axes.

Exciton density calculation

To calculate the density of interlayer excitons, we assume that the detected intensity per second $I$ is proportional to the density $n$ by the following expression: $I = n/\tau \cdot A \cdot \eta$, where $\tau$ is the decay time and $A$ the effective area of the detection spot ($A \approx \pi(d/2)^2$) for the confocal setup I with a diffraction limited spot of $d \sim 0.5\ \mu m$. The efficiency $\eta$ to detect a photon emitted from the sample is given by $\eta = \eta_{\text{Objective}} \cdot \eta_{\text{Optics}} \cdot \eta_{\text{Spectrometer}}$, with $\eta_{\text{Objective}}$ the collection efficiency of the objective with NA = 0.75). Following ref.(1), $\eta_{\text{Objective}}$ can be calculated to be

$$\eta_{\text{Objective}} = \frac{1}{2}\left[1 - \sqrt{1 - \frac{NA^2}{n^2}}\right] \cdot T = 0.019$$, where $n = 2.2$ is the refractive index of hBN and $T = 0.64$ is the transmission coefficient of the objective at a wavelength of 900 nm. $\eta_{\text{Optics}}$ is the accumulated efficiency of all optical elements between objective and spectrometer. From the specifications, we find $\eta_{\text{Optics}} = 0.007$. $\eta_{\text{Spectrometer}}$ is measured by illuminating the spectrometer with a cw laser at a wavelength of 900 nm with identical optics. The power is set to 0.3 nW, thus giving a number of incoming photons per second of $3.86 \cdot 10^6$. Acquiring a spectrum with an integration time of 1 second gives a total count number of $2.41 \cdot 10^5$. This leads to $\eta_{\text{Spectrometer}} = 0.062$. Therefore, the total efficiency of detecting a photon from the sample is $\eta = 8 \cdot 10^{-6}$. This value assumes perfect alignment and can only count as an upper limit to the real efficiency. We assume a lower bound of $4 \cdot 10^{-6}$ considering possible smaller efficiencies in the overall optical alignment. In cw excitation the density can be calculated by $n = I \cdot \tau/(A \cdot \eta)$. For pulsed excitation, the equation modifies to $n = \dfrac{I}{\left[\frac{T_{\text{Laser}}}{\tau}+1\right] \cdot f \cdot A \cdot \eta}$, with the repetition frequency $f$ and the pulse duration $T_{\text{Laser}}$.

Measurement of temporal coherence

The first order correlation function $g^1(t)$ is measured by Michelson-Morley interferometry. We use a beam splitter to split the emitted signal into two paths with equal intensity. Both paths exhibit a delay line. Path 1 introduces a rough delay $\Delta l$. The fine delay in path 2 is used to scan the interference locally around $\Delta l$. The resulting signal $I(\Delta l + l)$ is fitted by $I(\Delta l + l) = o + A \cdot \sin^2(\omega l + \varphi)$. ($o$: offset, : amplitude, $\omega$: frequency, $\varphi$: phase offset). The first order correlation function (visibility) is then calculated by $g^1(\Delta l) = A/(A + 2o)$. To exclude the impact of laser-induced coherence, we present the coherence visibility only for the off-period of the utilized pulsed laser.

Further, the signal is analyzed in a time-window of 50 ns exactly 50 ns after the laser pulse to exclude contributions from peaks (2) and (3), which have a decay-time well below 50 ns. We fit $g^1(\Delta l)$ with an exponential decay $A \cdot e^{-|\Delta l|/l_c}$ to extract the coherence length $l_c$. The coherence time is calculated by $\tau_c = l_c/c$, with the speed of light $c$.

Determination of the relative occupation numbers

At low excitation power and temperatures, the time-resolved photoluminescence of interlayer exciton PL shows a mono-exponential decay. Here, we conclude that only one type of interlayer exciton is present and the occupation is 100%. At higher excitation powers and temperatures a second decay channel appears. The resulting histogram can be modeled by

$$I(t) = \theta(t_0 - t) \cdot c + \theta(t - t_0) \cdot \frac{n_1^*}{\tau_1} \cdot e^{-\frac{t}{\tau_1}} + \theta(t - t_0) \cdot \frac{n_2^*}{\tau_2} \cdot e^{-t/\tau_2}.$$

The first term corresponds to the steady state during the laser pulse, during which the measured signal is constant: $I = c$. When the laser is off after $t = t_0$, the population decays exponentially within two channels of different decay times $\tau_1$ and $\tau_2$. The fit paramteres $n_1^*$ and $n_2^*$ are directly proportional to the exciton densities $n_1$ and $n_2$. The relative occupation $\xi$ can be calculated via $\xi = n_1^*/(n_1^* + n_2^*)$. The decay time of the laser after $t = t_0$ is measured to be $\tau_{\text{Laser}} = (0.64 \pm 0.09)$ ns.

Determination of diffusion lengths and coefficients

We excite interlayer excitons by a Gaussian shaped laser spot with a resolution of $\sigma_{\text{ex}} = 0.33$ μm (setup II). A spatial filter with a resolution of $\sigma_{\text{sf}} = 0.45$ μm in the detection path allows to spectrally and locally resolve the emission signal. We scan the detection spot across the sample by a dual-axis galvanometer mirror system in steps of 0.25 μm. The spectra at each position allow to extract the spatial extension of peaks (1), (2) an (3) individually. The extracted distributions can be fitted by a Gaussian function with a width of $\sigma_{\text{PL}(i)}$ ($i = 1,2,3$). To get the real distribution $\sigma_{\text{exciton}(i)}$ ($i = 1,2,3$) we deconvolve with the known resolution $\sigma_{\text{sf}}$. $\sigma_{\text{exciton}(i)} = \sqrt{\sigma_{\text{PL}(i)}^2 - \sigma_{sf}^2}$

We simulate the theoretical exciton density distribution for the given excitation profile $\Lambda \propto \exp(-r^2/2\sigma_{\text{ex}})$ and $0 = \frac{\partial n}{\partial t} = \Lambda - \frac{n}{\tau} - D \cdot \Delta n$, with the exciton density $n$, the decay time $\tau$ and the diffusion coefficient $D$. The relevant parameter describing the expansion is the characteristic diffusion length $L_D = \sqrt{D\tau}$. By comparing the simulated distribution $\sigma_{\text{exciton}}^{\text{theo}}$ with the experimental value $\sigma_{\text{exciton}(i)}$ we extract the diffusion lengths for each peak. By measuring the decay times $\tau_1$, $\tau_2$, and $\tau_3$ the diffusion coefficients $D_{(1)}$, $D_{(2)}$, and $D_{(3)}$ of peaks (1), (2), and (3) can be calculated.

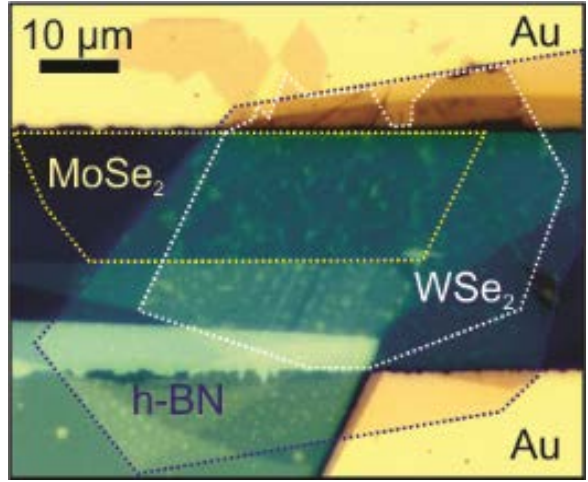

**Fig. S1. Microscope image of sample 1.**

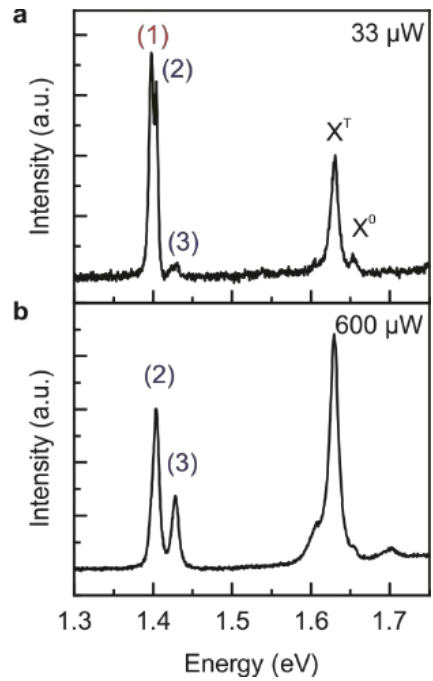

**Fig. S2. Photoluminescence spectra of intra- and interlayer excitons** at excitation powers of (**a**) 33 μW and (**b**) 600 μW ($T$ = 4K, cw excitation, $E_{Laser}$ = 2.541 eV). The energy difference between peaks (2) und (1) for $P_{Laser}$ = 33 μW is $\Delta E_{(2)-(1)}$ = (6.48 ± 0.08) meV.

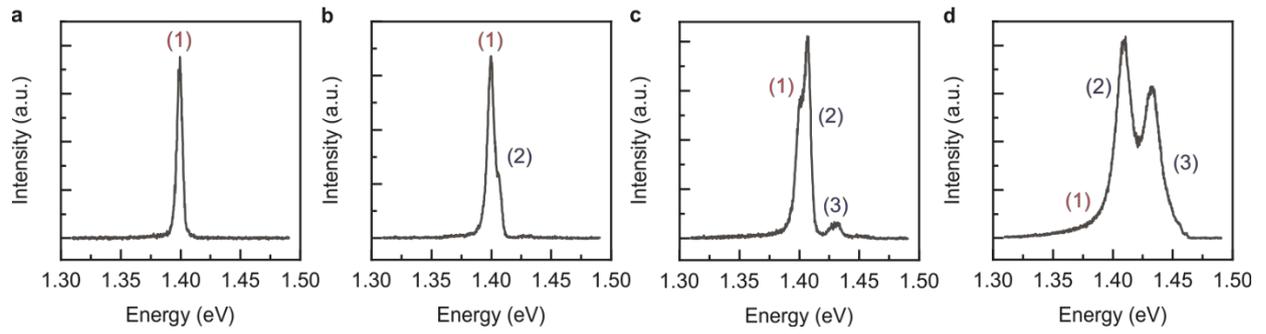

**Fig. S3. Photoluminescence spectra as a function of laser power. a,** At low excitation power: just peak (1) emits light (excitation power: 200 nW), **b,** above 5 µW, peak (2) appears on top of peak (1). **c,d,** at high powers, peaks (2) and (3) dominate [(**c**) 23 µW, (**d**) 420 µW]. Excitation energy $E_{photon}$ = 2.54 eV and $T$ = 4 K.

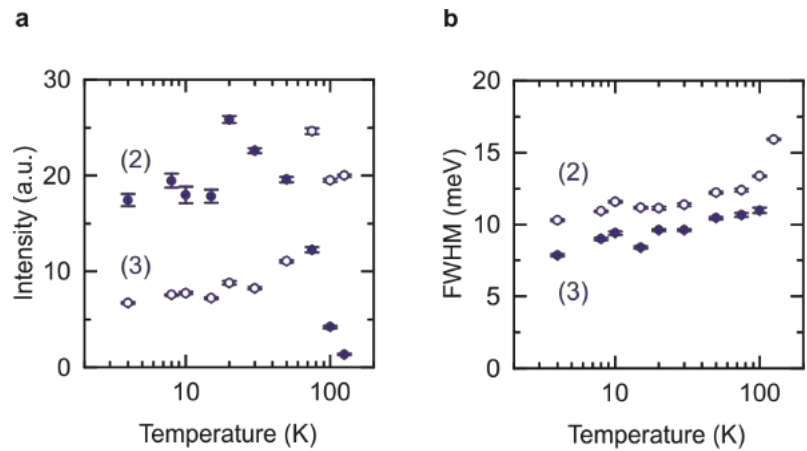

**Fig. S4. a,** Intensity and **b,** FWHM of peaks (2) and (3) vs. temperature (cw excitation, $E_{Photon}$ = 2.541 eV, $P_{Laser}$ = 600 µW).

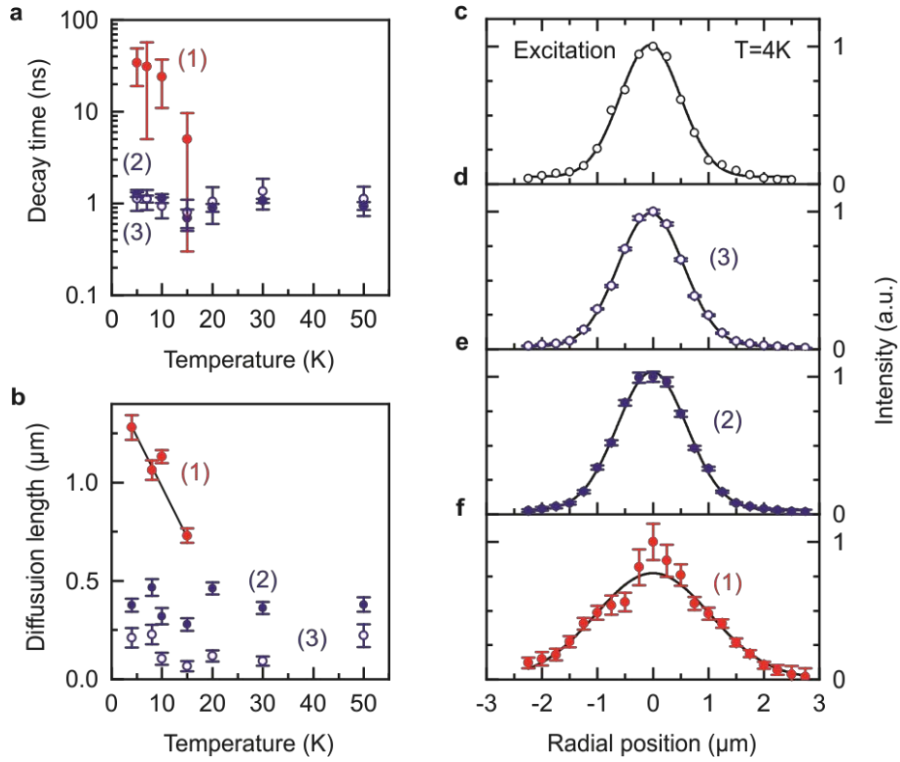

**Fig. S5. a, and b,** Decay times and diffusion length determined for peaks (1), (2), and (3) vs. temperature at an excitation power of $P_{Laser}$ = 25 µW ((a) $E_{photon}$ = 1.946 eV, pulsed with 200ns pulse length 25 µW, and (b) $E_{photon}$ = 2.541 eV, cw, 600 µW). The laser pulse duration is set to 200 ns at a repetition frequency of 500 kHz. **c,** Lateral scan across the excitation spot. The observed profile is a convolution of the excitation spot with the detection resolution ($\sigma_{sf}$ = 0.45 µm). **d,-f,** Lateral cross section of peaks (1), (2) and (3) at $T$ = 4K. The interlayer excitons are generated by a Gaussian laser spot (compare (c) with $\sigma$ = 0.33 µm, cw excitation, $E_{photon}$ = 2.541 eV, $P_{Laser}$ = 600 µW). A spatial filter is used to spectrally resolve the radial distribution of the peaks individually. **d,-f,** Spatial distribution of (**d**) peak (3), (**e**) peak (2), and (**f**) peak (1).

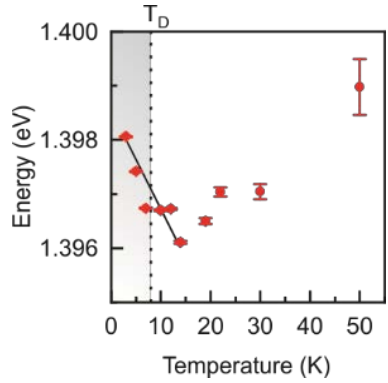

**Fig. S6.** Emission energy of peak (1) vs. temperature ($E_{photon}$ = 1.59 eV). These data are from the same set of measurements as Fig. 1c of the main manuscript. Line is a guide to the eye to highlight the increase of emission energy at low temperature, which is consistent with the dipole-dipole repulsion between interlayer excitons.

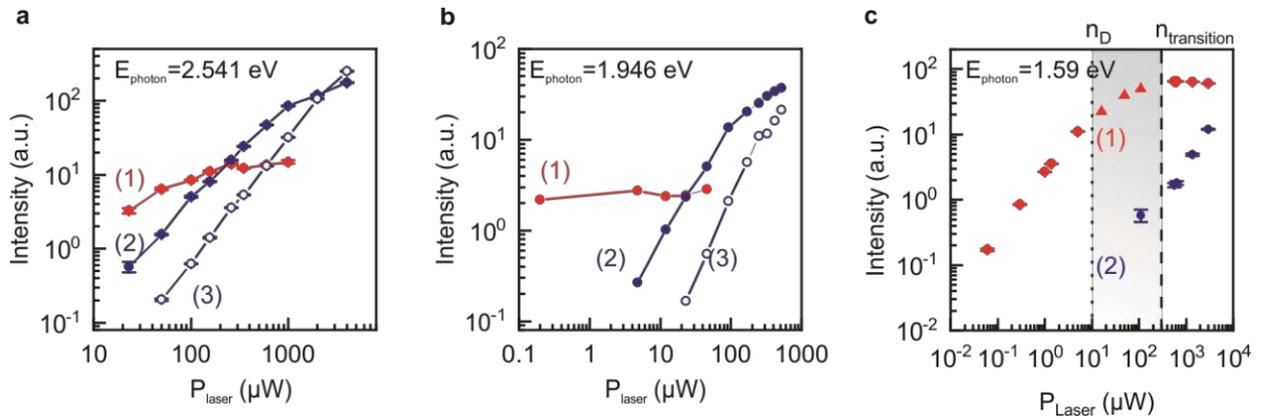

**Fig. S7.** Intensities of peaks (1),(2) and (3) vs. excitation power for different excitation energies. **a,** $E_{photon}$ = 2.541 eV. **b,** $E_{photon}$ = 1.946 eV **c,** $E_{photon}$ = 1.59 eV. For $E_{photon}$ = 1.59 eV, the set of data corresponds to the presented set of data in Fig. 3 of the main text. In turn, $n_D$ and $n_{transition}$ correspond to the density values as in Fig. 3 of the main text.

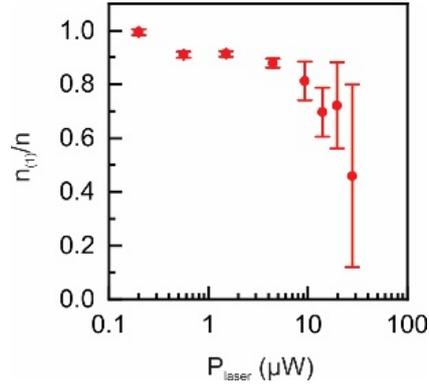

**Fig. S8. Relative occupation of peak (1) vs. excitation power** ($E_{photon}$ = 1.946 eV, $T$ = 4 K). The laser pulse duration is set to 200 ns at a repetition frequency of 500 kHz.

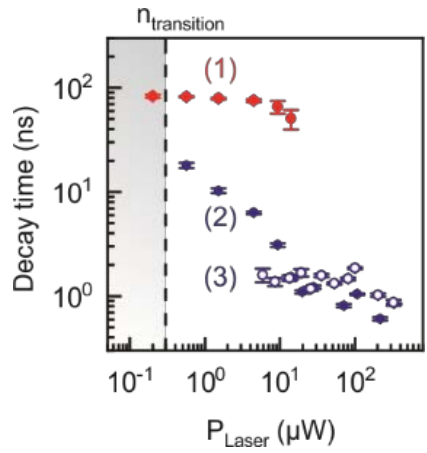

**Fig. S9.** Lifetimes $\tau_1$, $\tau_2$, and $\tau_3$ for peaks (1), (2), and (3) vs. excitation power at excitation energy $E_{photon}$ = 1.95 eV at $T$ = 4 K. For an excitation power below $n_{transition}$, only peak (1) emits light.

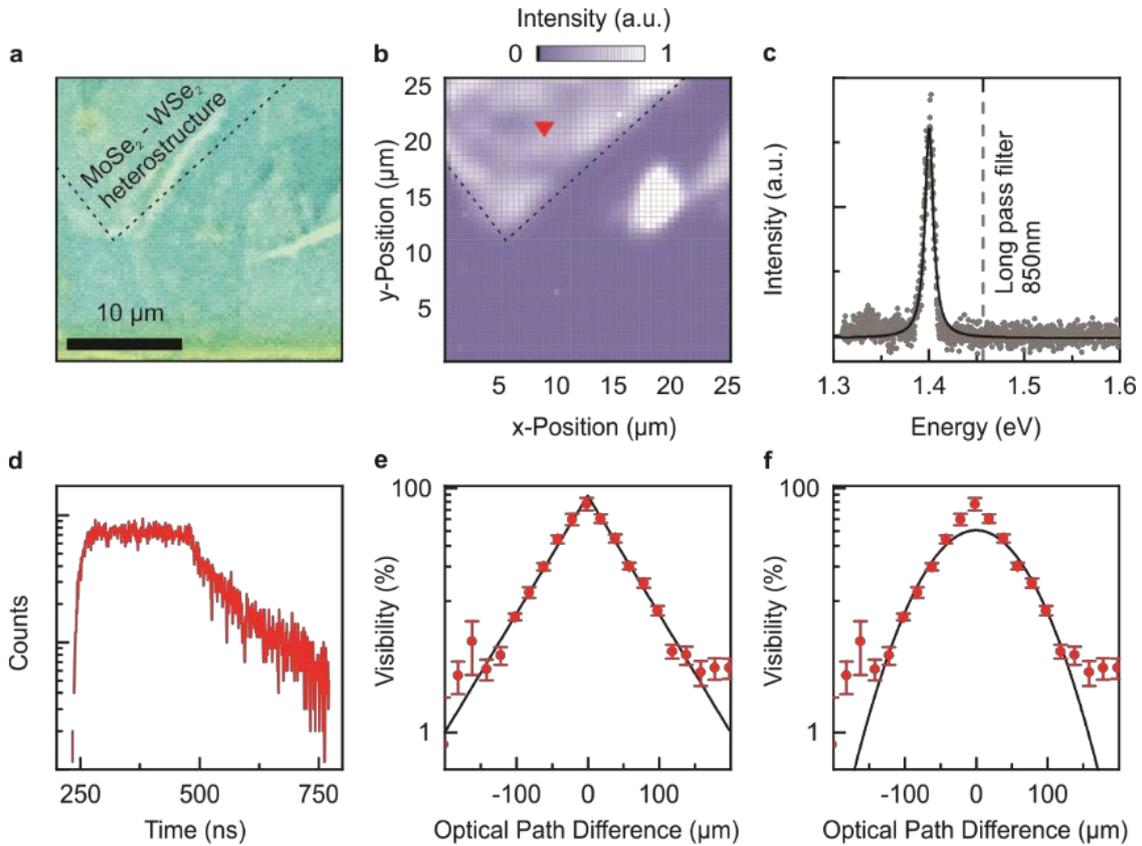

**Fig. S10. Temporal coherence of peak (1) on sample 2. a,** Optical micrograph of a MoSe$_2$-WSe$_2$ heterostack on a second device. **b,** Photoluminescence map at emission energy of peak (1) of the section in (A) at $T$ = 3.5 K, $P_{Laser}$ = 3 µW pulsed excitation, pulse duration 200ns, repetition frequency 500 kHz. **c,** Photoluminescence spectrum taken at position marked by triangle in (B). We use a 850nm long pass filter to suppress any signal above the interlayer transitions. **d,** Temporal photoluminescence signal with signal detection as in (c). The exponential decay time of peak (1) is $\tau_{(1)}$ = 78 ± 1 ns. **e,-f,** Temporal coherence measured via Michelson-Morley interferometry. The resulting visibility vs. optical path is fitted by (e) an exponential decay and (f) a Gaussian line shape. The corresponding total fit error in (f) is almost three times larger than in (e).

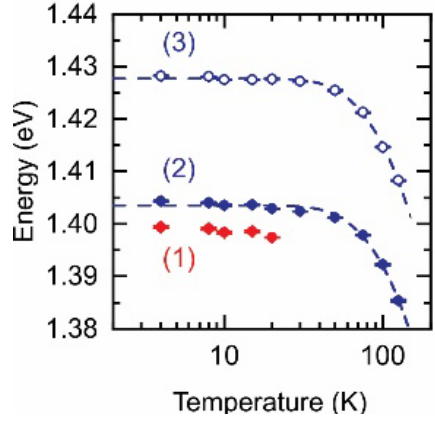

**Fig. S11. Emission energies of interlayer excitons [emission peaks (1), (2), and (3)] vs. temperature.** The temperature dependence of peak (2) and (3) can be described by a polaron shift (blue dashed lines). In this measurement, peak (1) could only be observed for temperatures up to 20 K. The temperature dependent emission energies of peaks (2) and (3) can be described phenomenologically by (1) $E_G(T) = E_G(0) - S\langle\hbar\omega\rangle\left[\coth\frac{\langle\hbar\omega\rangle}{2k_BT} - 1\right]$,[2] with $E_G(0)$ the peak position at $T = 0$ K, the Huang-Rhys factor $S$ and the average phonon energy $\langle\hbar\omega\rangle$. By the analysis, we get the following values for peak (2) [peak (3)] $E_G(0)$ = 1.404 ± 0.001 eV [1.428 ± 0.001 eV], S= 1.70 ± 0.21 [1.85 ± 0.11], and Ep = $\langle\hbar\omega\rangle$ = 14 ± 2 meV [14 ± 1 meV].

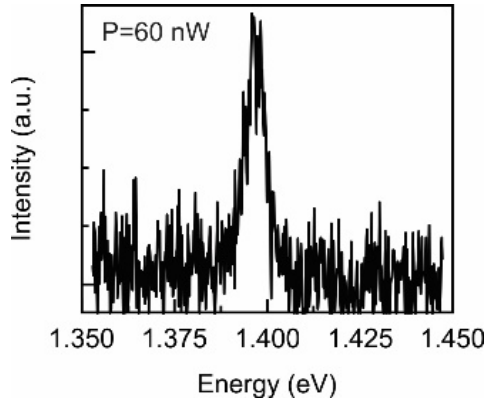

**Fig. S12. Emission spectrum of peak (1) at a low excitation power of $P_{Laser}$ = 60 nW** ($E_{photon}$ = 1.946 eV). The laser pulse duration is set to 200 ns at a repetition frequency of 500 kHz.

**Movie S1.** Spontaneous Synchronization. The movie S1 shows a classical example of a spontaneous synchronization. As soon as metronomes are put onto a movable blank, they start to interact with each other until all oscillators have the very same frequency. Similarly, below the degeneracy temperature, in the many-body regime, the investigated interlayer excitons start to interact via a dipole-dipole interaction until they unify to one state indicated by one emission energy and a Fourier-limited temporal coherence.

**References for the Supplementary Materials:**